\documentstyle [preprint,aps]{revtex}

\begin {document}

\draft

\title{Kinetic roughening model with opposite KPZ nonlinearities}

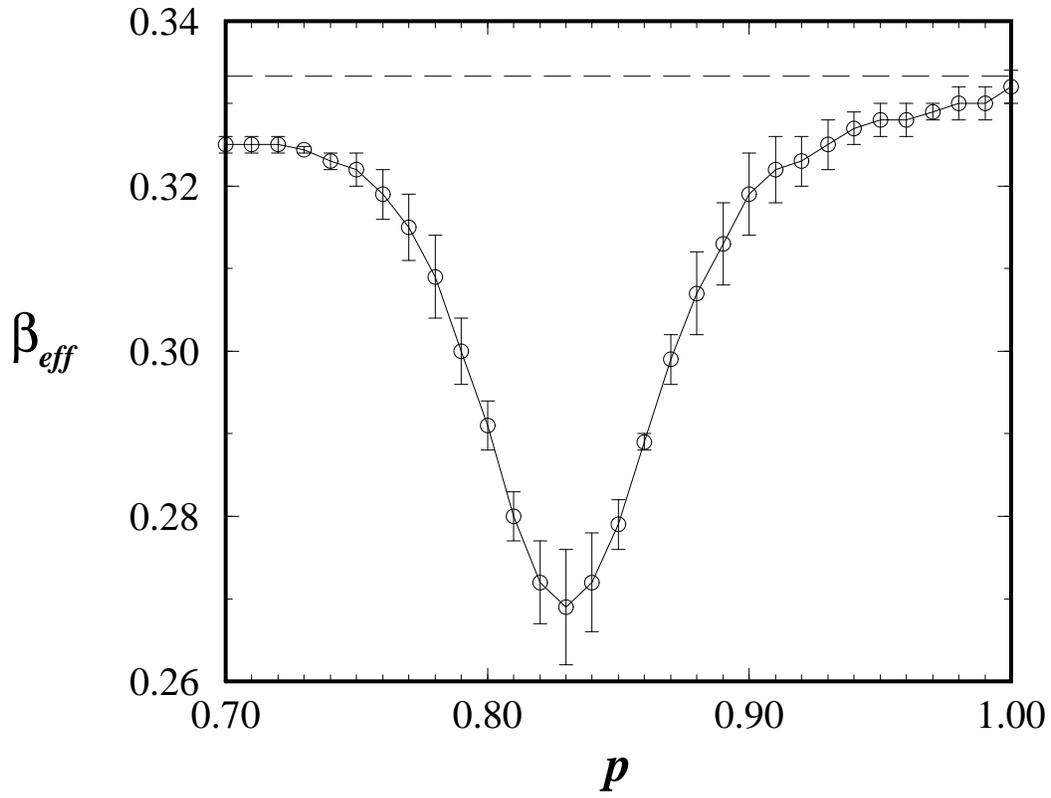
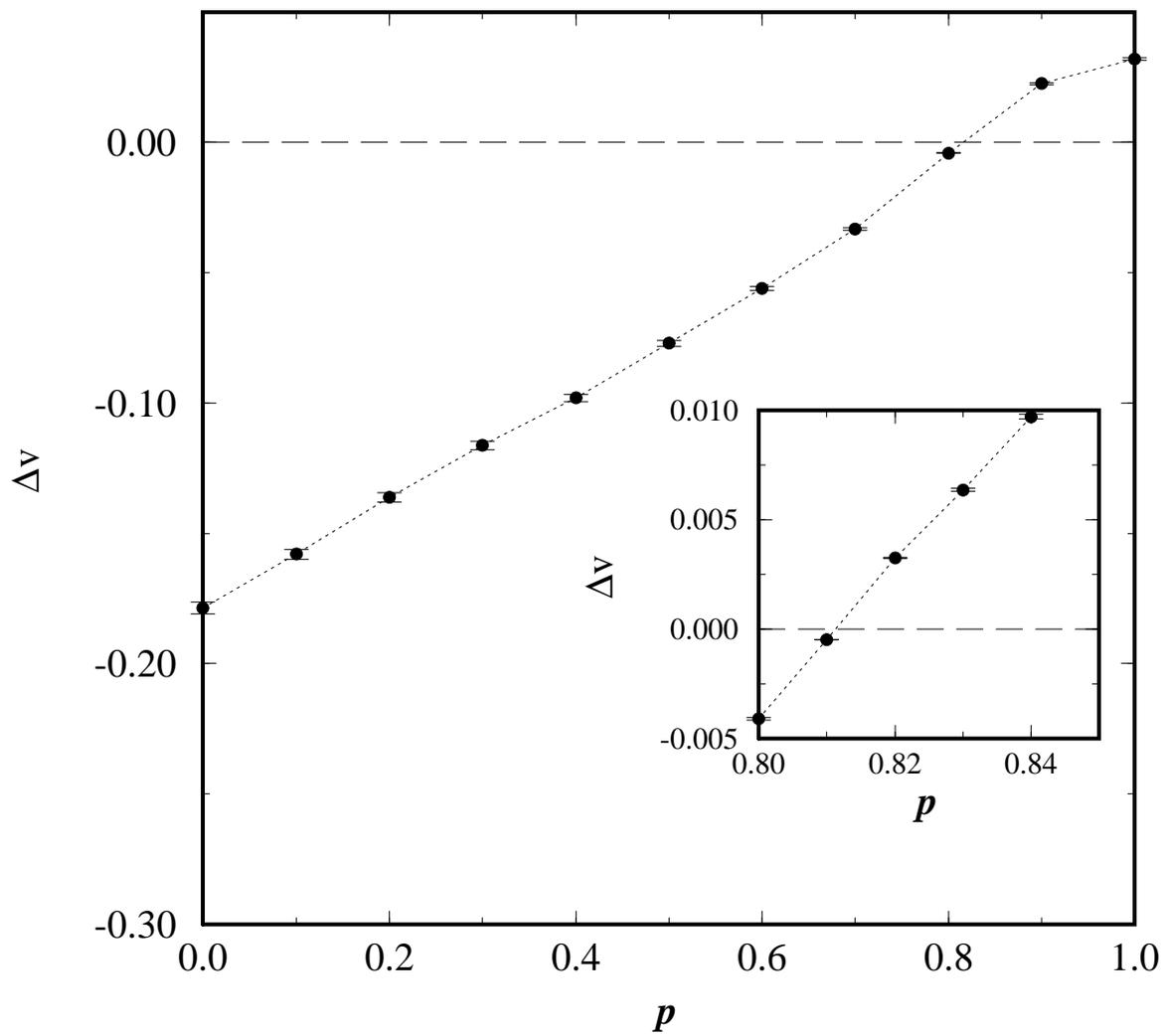
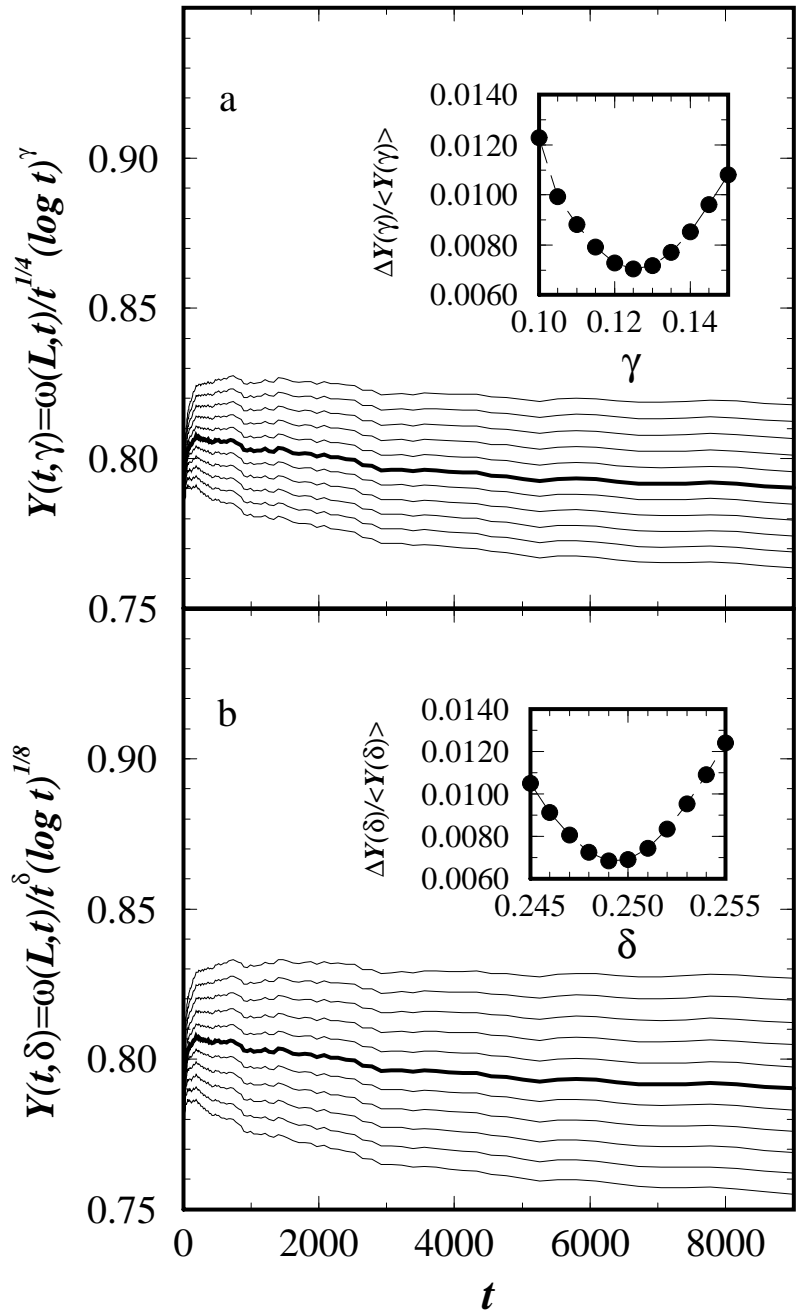
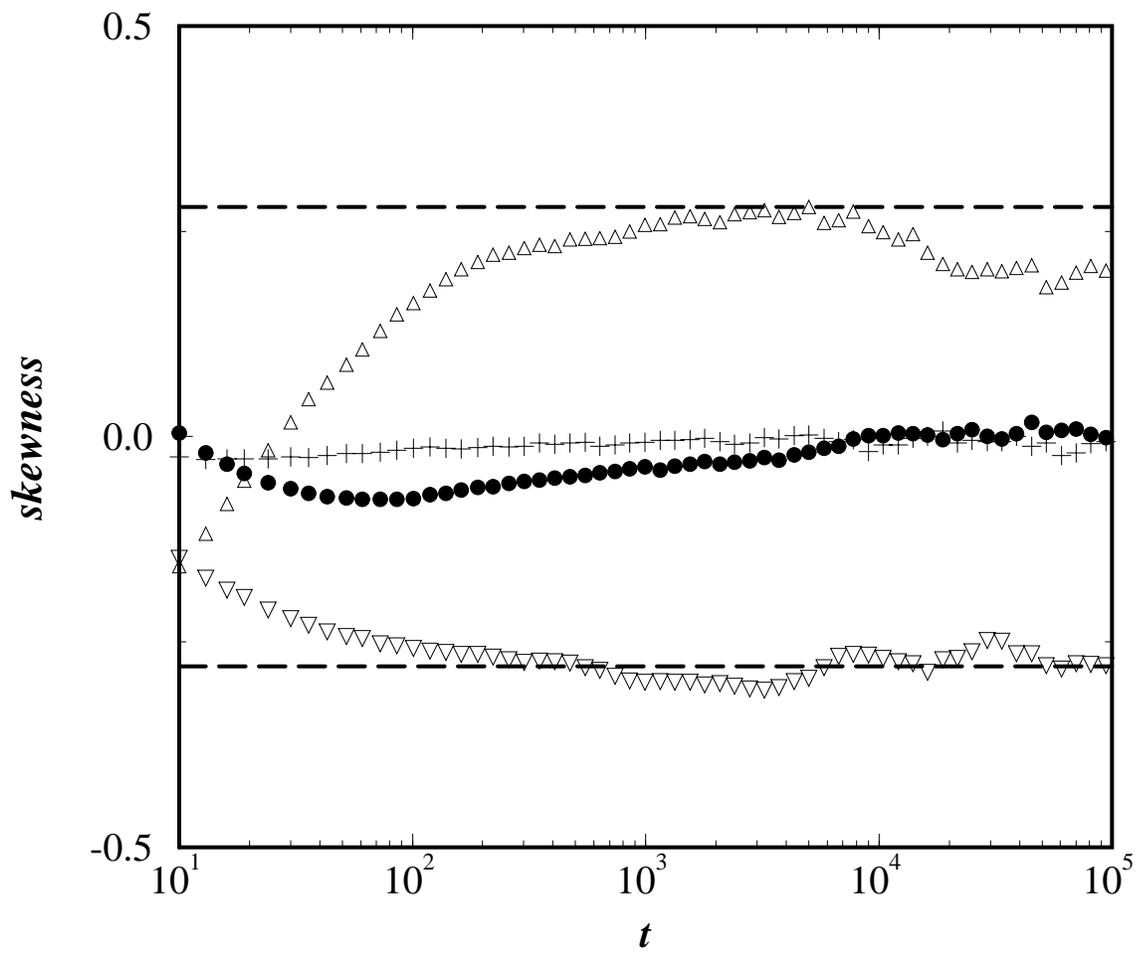
\author {T. J. da Silva and J. G. Moreira } 
\address {Departamento de F\'{\i}sica, Instituto de Ci\^encias Exatas,\\
Universidade Federal de Minas Gerais, C. P. 702,\\ 
30161-970, Belo Horizonte, MG - Brazil} 

\date {\today}
\maketitle

\begin {abstract}
We introduce a model that simulates a kinetic roughening process with two
kinds of particles: one follows the ballistic deposition (BD) kinetic and,
the other, the restricted solid-on-solid (KK) kinetic. Both of these
kinetics are in the universality class of the nonlinear KPZ equation, but
the BD kinetic has a positive nonlinear constant while the KK kinetic has
a negative one. In our model, called BD-KK model, we assign the probabilities
$p$ and $(1-p)$ to the KK and BD kinetics, respectively.
For a specific value of $p$, the system behaves as a quasi linear model and
the up-down symmetry is recuperated.
We
show that nonlinearities of odd-order are relevant in these low nonlinear
limit.
\end {abstract}
\vskip 2truecm


%\narrowtext 
%\twocolumn[\hsize\textwidth\columnwidth\hsize\csname @twocolumnfalse\endcsname
%\vskip2pc]
\newpage
%
%
%
%    --------------------- begin --------------------------------------
%
%
%
The growth of interfaces by nonequilibrium kinetic roughening models is
a very interesting topic of the far from equilibrium statistical mechanics
[1-3]. In the two past decades, several models have been proposed, as ballistic
deposition \cite{db}, Eden model \cite{eden}, solid-on-solid (SOS) model with
surface relaxation
\cite{ew,family86}, and SOS model with refuse \cite{kk}, SOS model with diffusion
\cite{wolfvillain}. In computer simulations, interfaces are described by
a discrete set $\{h_i(t)\}$ which represents the height of a site 
$i$ at the time $t$. Such interface has $L^d$ sites, where 
$L$ is the linear size and $d$ is the dimensionality of the substrate.
The
roughness of the interface is defined by
\begin{equation}
\omega^2(L,t)=\left<{1\over L^d}\sum_{i=1}^{L^d}(h_i-\bar h)^2\right>~~,
\end{equation}
where $\bar h$ is the mean height at the time $t$ and the symbols
$<..>$ means average over independent computational samples.
In most of the kinetic
roughening models, the roughness obeys the Family-Vicsek dynamic scaling
\cite{fv}
\begin{equation}
\omega(L,t)\sim L^{\alpha}f\left({t\over L^z}\right)~~,
\end{equation}
where the function $f(x)$ must be $L$-independent. 
The roughness behaves as $\omega\sim t^{\beta}$,
for short times $(1\ll t\ll L^z)$ and behaves as 
$\omega_{\infty}(L)\sim L^{\alpha}$ in the steady state.
The
$\beta$ and $\alpha$ are the growth and roughness exponents,
respectively, and are related with the dynamical exponent $z$
through the relation $z=\alpha/\beta$. For some systems, $\alpha=\beta=0$
and $z\neq 0$ which means that the roughness does not obey Eq.2 and
has a logarithmic behavior in space and time.

The kinetic growth models are also described by the continuum Langevin-like
equations in the coarse-grained limit. These equations
have terms which represents the main interactions
among the incoming particles. Example of a linear equation is 
the EW equation \cite{ew},
\begin{equation}
{\partial h({\bf x},t)\over\partial t}=v_0+\eta({\bf x},t)+
\nu\nabla^{2}h({\bf x},t)~~,
\end{equation}
that describes the fluctuations of the SOS model with surface
relaxation \cite{family86}. In Eq.3, the first two terms of the right side
are related to the deposition of particles. This deposition has a rate
$v_0$ and a noise with zero mean and variance given by
\begin{equation}
\left<\eta({\bf x},t)\eta({\bf x}^{'},t^{'})\right>=
 D\delta^{d}({\bf x} -{\bf x}^{'})\delta(t-t^{'})~~.
\end{equation}
\noindent The third term represents the surface relaxation process.

The exponents of Eq.3, obtained by Fourier analysis \cite{ew,nattang},
are $\beta(d)=(2-d)/4$, $\alpha(d)=(2-d)/2$ and $z(d)=2$.
For $d=1$, these expressions
gives $\beta=1/4$ and $\alpha=1/2$. For $d=2$, the scaling
exponents are $\beta=\alpha=0$.

There are also some nonlinear equations that describes nonlinear 
kinetic roughening models.
The more known nonlinear equation is the KPZ equation \cite{kpz}
\begin{equation}
{\partial h({\bf x},t)\over\partial t}=v_0+\eta({\bf x},t)+
          \nu\nabla^{2}h({\bf x},t)+{
          \lambda\over 2}\left(\nabla h({\bf x},t)\right)^{2}~~.
		  \end{equation}
This equation is more complete than Eq.4 because the nonlinear term may
represent the lateral growth, or the appearance of a driven force.
In $d=1$, the exponents of this equation\cite{kpz} are 
$\beta=1/3$, $\alpha=1/2$ and $z=3/2$. In $d=2$, the analytical solution is not
known. Examples of models in the universality class of KPZ equation are the
ballistic deposition (BD) \cite{db} and the SOS model with refuse (KK) \cite{kk}.
In $d=1$, numerical simulations \cite{texto} indicate $\beta\approx 0.30$,
$\alpha\approx 0.47$ for the BD model and
$\beta\approx 0.332$, $\alpha\approx 0.489$ for the KK model.
In $d=2$, these exponents are $\beta\approx 0.24$, $\alpha\approx 0.40$ (BD) 
and $\beta\approx 0.25$, $\alpha\approx 0.40$ (KK).

In this letter we report on results of computer simulation of a 
growth model with two kinds of particles. Both of them obeys the KPZ kinetic,
bu they have opposite signs of the nonlinear $\lambda$ constant.
The effort to understand the nonlinearity in stochastic
systems out of equilibrium is due to the great influence of the nonlinearities
in the scaling analisys, as pointed by Binder {\it et al}\cite{binder}.
The anisotropic KPZ equation, studied analytically in 1991 by Wolf
\cite {wolf1}, also contains two nonlinear terms with opposite
signs that describes the distinguishable directions of the substrate
in a vicinal surface.
He has found, by renormalization group, a logarithmic behavior of the
roughness ($\alpha=\beta=0$) for opposite signs of nonlinearity and
power-law behavior when the nonlinearities have same signs. This
results were confirmed by Kim {\it et al.} \cite{hjkim}
by simulations of a discrete model with two different kinetics applied in
each direction of the substrate.

The motivation of our work is the paper published by
Bernardes {\it et al} \cite{atb}. They have
studied  the deposition of particles with different radii on a cold substrate by
Monte Carlo simulation. The authors have found the growth
exponent $\beta\approx 0.26$, for $d=1$. So, they have concluded that the universality
class of this model is next to the EW class \cite{ew}. However, there are
two nonlinear characteristics in the morphology of the
model of Bernardes {\it et al}: (i) existence of
porosity in the bulk; (ii) the growth velocity, that is, 
$v=\left<{d{\bar h}/dt}\right>$, is greater than the deposition rate. 
These characteristics can indicate that the up-down symmetry
$(h\rightarrow -h)$ is broken \cite{barab}. 
The break of this symmetry leads to appearance of the nonlinear term in
the KPZ equation. Bernardes {\it et al.} also have checked the need of
logarithmic correction in the behavior of roughness \cite{binder,maya} that
indicates the presence of odd nonlinear terms in the growth equation.

The aim to create a model with opposite KPZ nonlinearities is to verify the
possibility of generating a linear process (when the effective
nonlinearity vanishes) with a morphology of a nonlinear growth process.
Our model is a probabilistic combination among the ballistic deposition model
(BD) and the SOS model with refuse (KK). The KK model occurs with
$p$-probability, and the BD model occurs with $(1-p)$-probability.

Ballistic deposition is a process where particles are
dropped vertically onto a smooth substrate. The incoming particles are
automatically joined to the growing cluster when its first contact with the
growing interface occurs. In the in-lattice version, we
select at random a site $i$ of the lattice
and its new height is evaluated by the algorithm \cite{meakim1}
\begin{equation}
h_i'=max(h_i+1;h_{\{j\}})~~,
\end{equation}
where $\{j\}$ are the first neighbors of the site $i$. The BD kinetic
does not generate a solid-on-solid deposition because it generates a
structure with porosity. Therefore, we define
the growing profile as the major height of an occupied site of each column.

The KK model is a SOS random deposition with the difference of
height constraint $h_i-h_{\{j\}}<m$,
where $m$ is the parameter that controls the roughness.
If the height of the particle deposited on the site $i$ 
did not satisfy the height constraint, this particle is not incorporated to
the interface.

Both of them are in the same universality class of KPZ equation. We have 
chosen these two models because: (i) the BD model generates a bulk with
porosity and has nonlinear parameter $\lambda_{BD}>0$ because the growth
velocity is bigger than the deposition rate $v_0$; (ii) the KK model
has $\lambda_{KK}<0$ because the kinetic of refuse makes the growth velocity
smaller than the rate of deposition.

In our simulations, a unit of time means that we have done $L$ attempts of deposition.
Moreover, all simulations were done with a one-dimensional substrate (d=1) and, in
the KK model, the difference of height constraint $m=1$.

Figure 1 shows the plot of the effective growth exponent $\beta_{eff}$ {\it vs.} the
parameter $p$, for
$L=50,000$. The long dashed line is the exact value of $\beta$ obtained from
KPZ equation by renormalization group.
We have done 100 independent runs for each probability and we applied 
consecutive slopes method \cite{barab}
in the log-log plots of $\omega$ {\it vs.} $t$ 
in the time interval $20<t<10000$, giving an ensemble of 
$\beta_{eff}$
exponents for each value of $p$. So, we
have estimated the error bars around each value of $\beta_{eff}$.
For $p^*=0.83$, the model is near to the EW class,
because the growth exponent is $\beta_{eff}\approx 0.27$. However, the error
bars in this region has increased, indicating the need of
more careful analysis of the scaling.
These results can point that the effective KPZ term was
removed in the probability $p^*$.

In order to make better characterization of the KPZ nonlinearity in
our model, we
do finite size analysis in the growth velocity. In 1990, Krug and Meakin
\cite {krugmeakim} showed that the finite size correction,
$\Delta v(L,t)=v(L,t)-v_{\infty}$, for a model in KPZ class behaves as
\begin {equation}
\Delta v(L)\sim -\lambda L^{-\alpha_{\parallel}}~~,~~{\rm for}~~t\gg L^z~~,
\end {equation}
where the $\alpha_{\parallel}$ exponent depends on the roughness exponent.
The $\Delta v$ correction goes to zero when the KPZ term
vanishes. So, with Eq.7, we can obtain the sign of the KPZ
nonlinearity and determine when the nonlinearity goes to zero in function
of the tunning parameter $p$. 
Figure 2 shows the plot of $\Delta v=v(L=10)-v(L=1280)$ {\it vs.} $p$ for
the BD-KK model. The finite size correction vanishes for $p\approx 0.81$
(see inset),
very close to the value obtained in the minimum of the
$\beta$ {\it vs.} $p$ plot (see Figure 1). 

We also check the need of multiplicative logarithmic corrections in
the scaling for $p=p^*$. When the growth equation for a process has
a sequence of odd nonlinear terms as
\begin {equation}
{\partial h({\bf x},t)\over\partial t}=\eta(x,t)+
\nu{\partial^2 h({\bf x},t)\over\partial x^2}+\sum_{2n+1}
 \lambda_{2n+1}{\left({\partial h({\bf x},t)\over\partial x}\right)}^{2n+1}~~,
\end {equation}
for $n=1,2,...$, the roughness behaves as \cite{binder,maya}
\begin{equation}
\omega(L,t)\sim t^{1/4}(\log t)^{1/8}~~,
\end {equation}
for $t\ll L^z$.
So, if logarithmic corrections are accepted for $p=p^*$, it means that the
system is marginally in the EW class. 

In order to show that Eq.9 is really the better equation that describes
the system near to $p^*$, we rewrite this equation as
\begin {equation}
\omega(L,t)\sim t^{\delta}(\log t)^{\gamma}~~,
\end {equation}
and we do small variations around the exact values $\delta=1/4$
and $\gamma=1/8$. We analyze the validation of the Eq.10 by the
evaluation of the deviations from the horizontal curve
$Y(\delta,\gamma,t)=\omega(L,t)/t^{\delta}(\log t)^{\gamma}$ {\it vs.} $t$,
because this emphasizes better the deviations of the behavior
from this equation.
So, we measure the relative error $\Delta Y/\left<Y\right>$
of each curve $Y(\delta,\gamma,t)$ 
in function of the variations in $\delta$ and $\gamma$.
Figure 3a shows the plots of $Y(t,\gamma)$ {\it vs.} $t$ with $\delta=1/4$
and Figure 3b shows the plots of 
$Y(t,\delta)$ {\it vs.} $t$ with $\gamma=1/8$. The insets shows the
relative error $\Delta Y/\left<Y\right>$ {\it vs.} the scaling exponent
related to each case, $\gamma$ or $\delta$, respectively. 
The relative error, for the two cases, 
reaches a minimum when the exponents $\gamma=1/8$ and $\delta=1/4$, indicating
that Eq.9 is a good description for the temporal behavior of roughness.
This is pointing out that odd-nonlinear terms are relevant.

The nonlinear Eq.8 preserves the up-down symmetry, but this is not
obvious in BD-KK model at $p=p^*$, and also in the growth model of
Bernardes {\it et al.}. 
Eq.1 can be generalized for any moment $q$ of the height
distribution, in $d=1$, as
\begin{equation}
\omega^q(L,t)=\left<{1\over L}\sum_{i=1}^{L}(h_i-\bar h)^q\right>~~,
\end{equation}
and we concentrate on the behavior of odd-moments, especially the
third moment $(q=3)$ which is related to up-down symmetry. The skewness, 
defined by ${\cal S}=\omega^3/(\omega^{2})^{3/2}$,
can show us if the system has this symmetry. If the
skewness is null, the system has up-down symmetry because all odd moments
of the height
distribution vanish. On the other hand, for systems without the
up-down symmetry and in the KPZ class, some authors \cite{nijs1,nijs2}
believe that the skewness has
a universal value $|{\cal S}|\approx 0.28$.
Figure 4 shows the skewness ${\cal S}$ as a function of the time $t$ for
the model with positive nonlinearity ($p=0.0$, triangle up), 
with negative nonlinearity ($p=1.0$, triangle down)
and at the low nonlinear point $p=p^*$ (filled circles). 
As illustration, we also show the curve for
the model with surface relaxation \cite{family86} which is a linear
model and has the up-down symmetry (plus). We note that the skewness for $p=0$
tends to ${\cal S}=0.28$ and, for $p=1.0$, to ${\cal S}=-0.28$. For
$p=p^*$, in the asymptotic limit, the skewness goes to zero, suggesting the
presence of the up-down symmetry.

In conclusion, we have studied the scaling properties of a model with
opposite signs of the KPZ nonlinearity through numerical simulations
of a model with two kinds of particles. The Kim-Kosterlitz kinetic
occurs with a probability $p$ and the ballistic deposition model with a
probability $(1-p)$. For a specific value of the tunning parameter $p$, 
we show that the KPZ nonlinearity goes to zero
and the up-down symmetry is recuperated. 
We also show that odd-nonlinearities are 
relevant in this model. 

We would like to acknowledge Jos\'e Francisco de Sampaio for reading this
manuscript and Am\'erico T. Bernardes for fruitful discussions.
The numerical simulations were made in part in Alphas 500 Au of the
Departamento de F\'{\i}sica -UFMG and in part in a
Sun HPC 10000 of the Cenapad MG-CO. This work was supported by
CNPq, Fapemig and Finep/Pronex, Brazilian agencies.  
                                                                              
% 
%
%    -----------------------   end   --------------------------------
%
%
%

\newpage

{\large \bf Figure captions}

\bigskip

{\bf Figure 1}

The growth exponent $\beta$ {\it vs.} the parameter $p$ plot of the BD-KK model for $L=50,000$. 
The long-dashed
line is the exact value of the $\beta$ exponent for the KPZ equation.

{\bf Figure 2}

Plot of the difference between the steady state growth velocities
for $L=10$ and $L=1280$ {\it vs.} the parameter $p$ which gives the amount of KPZ
nonlinearity in the system. The inset is showing the behavior next to the crossover.

{\bf Figure 3}

Plots of the function $Y(t,\gamma,\delta)$ {\it vs.} the time $t$: (a) 
$\delta=1/4$ and small variations in $\gamma$ are performed,
(b) $\gamma=1/8$ and small variations
in $\delta$ are performed. The bold curves are showing the behavior of the
function $Y(t,1/8,1/4)$. The insets shows the
relative error $\Delta Y/\left<Y\right>$ {\it vs.} the scaling exponent
related to each case. 

{\bf Figure 4}

Plots of the skewness ${\cal S}$ as a function of the time $t$ for $p=0.0$ 
(triangle up),
$p=1.0$ (triangle down) and $p=p^*$ (filled circles). The skewness for the
EW linear model is represented by the symbol plus. The long dashed straight lines
indicate the $\pm 0.28$ estimated values. All simulations were done with $L=50,000$.

\end{document}